\documentclass[aps,amsmath,amssymb,prl,preprint,groupedaddress,
showpacs,twocolumn]{revtex4-1}
\usepackage{graphicx}
\usepackage{snapshot}
\begin{document}

\title{Multiplicative Propagation of Error During Recursive Wavelet Estimation}

\author{Michael A. Cohen}
\email[]{mike@cns.bu.edu}
\affiliation{Department of Cognitive and Neural Systems, Boston University}
\author{Can Ozan Tan}
\email[]{cotan@partners.org}
\affiliation{Cardiovascular Research Laboratory, Spaulding Rehabilitation 
Hospital and Department of Physical Medicine and Rehabilitation, Harvard 
Medical School}
\thanks{Corresponding Author. CVLab, Spaulding Rehabilitation Hospital, 
125 Nashua Street, Boston, MA 02114}
\date{\today}

\begin{abstract}
Wavelet coefficients are estimated recursively at progressively
coarser scales recursively. As a result, the estimation is prone to
multiplicative propagation of truncation errors due to quantization
and round-off at each stage. Yet, the influence of this propagation
on wavelet filter output has not been explored systematically. Through
numerical error analysis of a simple, generic sub-band coding scheme
with a half-band low pass finite impulse-response filter for down
sampling, we show that truncation error in estimated wavelet filter
coefficients can quickly reach unacceptable levels, and may render
the results unreliable especially at coarser scales. 
\end{abstract}


\maketitle

\section{Introduction}
With the exception of Haar wavelets, wavelet coefficients
(e.g., in Coiflets or Daubechies wavelets) are approximations to equations
without closed--form solutions. Coefficients are recursively estimated
at progressively coarser scales, using the estimate on one scale as
the input to the next. Such recursive schemes are prone to multiplicative
propagation of errors due to quantization and round-off ("truncation
errors") in estimation in each scale. Sub-band coding (i.e., breaking
the signal into a number of different frequency bands and encoding
each one independently) and downsampling at each scale may determine
the pattern of errors. Moreover, these truncation errors are not improved
by sample size, and can dominate the variance of estimation. Yet,
the propagation of truncation errors and their influence on estimated
wavelet coefficients have not been explored systematically. In this
note, we report the results from an error analysis of a simple, generic
sub-band coding scheme with a half-band low pass finite impulse-response
(FIR) filter for down sampling. We demonstrate the sub-band coding
scheme, and derive the bounds for signal-to-noise ratio of the filter
output in Section~\ref{sec:Sub-Band-Coding}, and provide a numerical
analysis of truncation error propagation in 
Section~\ref{sec:Numerical-Analysis}.

\section{Signal-to-Noise Ratio, Sub-Band Coding, and Downsampling
\label{sec:Sub-Band-Coding}}

\noindent Due to the dependencies between the variables and the complexity
of the expression, an exact formula for the signal-to-noise ratio
(SNR) is intractable in general. However, an approximate expression
(in decibels) can be given for the SNR

\noindent 
\begin{equation}
10\log_{10}\mathbb{E}\left[\dfrac{\mathcal{S}}{\mathcal{N}}\right]\label{eq:snr}
\end{equation}

\noindent where $\mathcal{S}$ and $\mathcal{N}$ denote the power
of the signal and noise. With small relative error, the expression
in brackets can be replaced with
$\dfrac{\overline{\mathcal{S}}}{\overline{\mathcal{N}}}$,
where $\bar{\cdot}$ denotes the mean. Assuming the respective coefficients
of variation, $\mathfrak{c}_{\mathcal{S}}=\dfrac{\sigma(\mathcal{S})}{\overline
{\mathcal{S}}}$,
and $\mathfrak{c}_{\mathcal{N}}=\dfrac{\sigma(\mathcal{N})}{\overline{\mathcal
{N}}}$
are small (i.e. bounded by a a small constant $\epsilon$), the minimal
relative error can be obtained as 

\begin{equation}
\dfrac{\mathcal{S}}{\mathcal{N}}-\dfrac{\overline{\mathcal{S}}}{\overline{
\mathcal{N}}}=\dfrac{\overline{\mathcal{S}}}{\overline{\mathcal{N}}}\left(
\dfrac{\left[\dfrac{\mathcal{S}-\overline{\mathcal{S}}}{\sigma_{\mathcal{S}}}
\right]\mathfrak{c}_{\mathcal{S}}-\left[\dfrac{\mathcal{N}-\overline{
\mathcal{N}}}{\sigma_{\mathcal{N}}}\right]\mathfrak{c}_{\mathcal{N}}}{1+
\left[\dfrac{\mathcal{N}-\overline{\mathcal{N}}}{\sigma_{\mathcal{N}}}\right]
\mathfrak{c}_{\mathcal{N}}}\right)
\end{equation}

\noindent after a simple algebraic manipulation. The term in parenthesis
consists of two approximately normally distributed random variables
(given a large sample size) with zero mean and a small variance, divided
by a normally distributed random variable with mean one and a small
variance. Therefore, the difference on the left-hand side will be
relatively small with high probability. From now on, we assume the
numerator and denominator in equation \ref{eq:snr} can be replaced
by its expectation. 

Let $f_{k}$ be the downsampling filter. Then, for a fixed input $S$
using a noise free filter of length $N$, the detail coefficients
$d_{yz}$ take the form

\begin{equation}
d_{yz}=S_{2^{z}y}-\sum_{i_{k}}\prod_{{k=1\atop i_{k}\in\{1:N\}}}^{z}f_{i_{k}}
S_{2^{z}y-\sum_{k=1}^{z}2^{k-1}i_{k}}
\end{equation}

\noindent When each of the filter coefficients $f_{i}$ is perturbed
by noise ($\epsilon_{i}\sim\mathcal{N}(0,\lambda_{i})$, where $\lambda_{i}$
is chosen to fix the input signal-to-noise ratio, $\mathrm{SNR}_{I}$)
, the perturbed detail coefficient, $\tilde{d}_{yz}$, takes the form

\begin{equation}
\tilde{d}_{yz}=d_{yz}=S_{2^{z}y}-\prod_{k=1}^{z}(f_{i_{k}}+\epsilon_{i_{k}})
S_{2^{z}y-\sum_{k=1}^{z}2^{k-1}i_{k}}\label{eq:perturbed_coef}
\end{equation}

\noindent At scale level $z$, the output signal to noise ratio 
$\left(\mathrm{SNR}_{O}\right)$ is then given (in decibels) by 

\begin{equation}
10\log_{10}\dfrac{\mathbb{E}\sum_{y}d_{yz}^{2}}{\mathbb{E}\sum_{y}(d_{yz}-
\tilde{d}_{yz})^{2}}\label{eq:SNR_O}
\end{equation}

\noindent where

\begin{widetext}
\begin{eqnarray}
d_{yz}-\tilde{d}{}_{yz} & = & \sum_{k,i_{k}}\epsilon_{i_{k}}\prod_{{j=1\atop j
\neq k}}^{z}(f_{i_{j}}+\epsilon_{i_{j}})S_{2^{z}y-\sum_{k=1}^{z}2^{k-1}i_{k}}
\nonumber \\
 & = & \sum_{k=1}^{z}\prod_{{j\in S_{k}^{z}\atop 1\leq i_{j}\leq N}}
\epsilon_{i_{j}}\prod_{{j'\in\{1:z-S_{k}^{z}\}\atop 1\leq i_{j'}\leq N}}
f_{i_{j'}}S_{2^{z}y-\sum_{k=1}^{z}2^{k-1}i_{k}}\label{eq:ndiff}
\end{eqnarray}
\end{widetext}

\noindent Assuming that $E[S_{i}S_{j}]=\delta_{ij}$, we can approximate
the numerator as

\begin{widetext} 
\begin{align}
\mathbb{E}(d_{yz}^{2})= & 1-\sum_{\{i_{k}|k=1\cdots
z\}}2f_{-\sum_{k=2}^{z}2^{k-1}i_{k}}\prod_{k=2}^{z-1}f_{i_{k}}\nonumber \\
 & +\sum_{\{i_{k,}i'_{k}|k=1\cdots
z\}}f_{i_{0}+\sum_{k=2}^{z}2^{k-1}(i_{k}-i'_{k})}\prod_{k=1}^{z}f_{i_{k}}\prod_{k=2}^{z}
f_{i'_{k}}
\end{align}
\end{widetext}

Note that the indices $\{i_{1},i_{k},i'_{k},k=2\cdots z\}$ are freely
varying in the above equation. Since $f_{k}$ is a low pass filter
with unity gain, we can assume $\sum_{k}f_{k}=$1. We also assume
that each of the $f_{k}$ are positive (there are many low pass filters
which are strictly positive, and the gain can always be adjusted).
Under these assumptions, and assuming no exclusions due to truncation,
it is easy to show that bounds for the numerator,
$\mathbb{E}\left(d_{yz}^{2}\right)$,
are

\begin{widetext}
\begin{equation}
\max(0,1-2\max_{k}f_{k}+\min_{k}f_{k})\leq\mathbb{E}(d_{yz}^{2})\leq1-2\min_{k}
f_{k}+\max_{k}f_{k}
\end{equation}

\begin{equation}
10\log_{10}\mathbb{E}(\sum_{y}d_{yz}^{2})\leq10\log_{10}N-10z\log_{10}2+\log_{10}
(1-2\min_{k}f_{k}+\max_{k}f_{k})\label{eq:numup}
\end{equation}

\begin{equation}
10\log_{10}\mathbb{E}(\sum_{y}d_{yz}^{2})\geq10\log_{10}N-z\log_{10}2+\log_{10}
(\max(0,1-2\max_{k}f_{k}+\min_{k}f_{k}))
\end{equation}
\end{widetext}

\noindent These inequalities will be true for real data with a very
small margin of error. However, only inequality \ref{eq:numup} is
likely to carry much weight.

Further note that by assumption,
$\mathbb{E}[\epsilon_{i}\epsilon_{j}]=10^{-\mathrm{SNR}_{I}/10}\dfrac{\left\Vert
S^{2}\right\Vert }{N}\delta_{ij}$.
Therefore, in general,

\begin{widetext}
\begin{align}
10\log_{10}\overline{\mathcal{N}} &
=\mathbb{E}\sum_{y}(d_{yz}-\tilde{d}_{yz})^{2}\nonumber \\
 & =10\log_{10}\sum_{j}(10^{-j(\mathrm{SNR}_{I}/10)}\dfrac{E\left\Vert S\right\Vert
^{2j}[2j]!}{N^{j}2^{j}j!}\mathfrak{f_{j}}\nonumber \\
 &
=-\mathrm{SNR}_{I}-10\log_{10}\mathfrak{f_{1}}+10\log_{10}\left[1+\sum_{j=1}^{K}
10^{-j(\mathrm{SNR}_{I}/10)}\dfrac{[2(j+1)]!\mathfrak{f_{j}}}{2^{j+1}j!
\mathfrak{f_{1}}}\right]\label{eq:filt}\\
 & \approxeq-\mathrm{SNR}_{I}-10\log_{10}\mathfrak{f_{1}}
\end{align}
\end{widetext}

\noindent where $\mathfrak{f}_{j}$ is a factor that depends only
on the filter coefficients $f_{j}$ under the above assumptions. Equation
\ref{eq:filt} follows since $\left\Vert S\right\Vert ^{2}=N,$ as
$S$ is assumed to be white noise. The expansion in equation \ref{eq:filt}
involves terms weighting the higher Gaussian moments involving the
perturbation of $f$ in equation \ref{eq:perturbed_coef}. 

The calculation of the expectation of the denominator in equation
\ref{eq:SNR_O} is quite complicated. However, if we further assume
that the input SNR is sufficiently small so that the summand in equation
\ref{eq:filt} is negligible, we only need to calculate $\mathfrak{f_{1}}$.
Under these assumptions, moments higher than the second in the matching
random variable need not be considered, and 

\begin{widetext}
\begin{equation}
\begin{array}{ccc}
\mathfrak{f_{1}} & =\sum_{y} & \mathbb{E}\left[\sum_{\{i_{j,},i_{j'}',j,j'=1
\cdots z\}}\epsilon_{i_{j}}\epsilon_{i'_{j}}\prod_{{k,k'=1\atop {k\neq j\atop k'\neq
j'}}}^{z}f_{i_{k'}}f_{i'_{k'}}S_{2^{z}y-\sum_{k=1}^{z}2^{k-1}i_{k}}S_{2^{z}y-
\sum_{k=1}^{z}2^{k-1}i'_{k'}}\right]\\
 & = & \sum_{y}\left[\sum_{\{i_{j,},i'_{j'},j,j'=1
\cdots z\}}\delta(i_{j}-i'_{j'})\prod_{{k,k'=1\atop {k\neq j\atop k'\neq 
j'}}}^{z}f_{i_{k'}}f_{i'_{k}}\delta(\sum_{k=1}^{z}2^{k-1}\delta(i_{k}-i'_{k'}))\right]
\\
 & = & \dfrac{N}{2^{z}}\left[\sum_{\{i_{j,},i'_{j'},j,j'=1
\cdots z\}}\prod_{{k,k'=1\atop {k\neq j\atop k'\neq
j'}}}^{z}f_{i_{k'}}f_{i'_{k}}\delta(\sum_{k=1}^{z}2^{k-1}(i_{k}-i_{k'}'))
\delta(i_{j}-i'_{j'})\right]
\end{array}\label{eq:f1}
\end{equation}
\end{widetext}

\noindent Equation \ref{eq:f1} shows that we have all products of
all filter coefficients of length $2z$, but two are excluded. The
excluded ones represent the same scale, and are required to be evaluated
at the same index. One remaining coefficient is fixed so that both
products of coefficients would contribute to the same detail coefficient.
To obtain a bound on this sum of products, we again assume $\sum_{j}f_{j}=1$.
In this case,
\[
N^{2}z/2^{z}\min|f|\leq\mathfrak{f}_{1}\leq N^{2}z/2^{z}\max f
\]

\noindent and thus,

\begin{widetext}
\begin{equation}
10\log_{10}\dfrac{\sum_{y}d_{yz}^{2}}{\mathbb{E}\sum_{y}(d_{yz}-d_{yz}^{p})^{2}}
\leq\mathrm{SNR}_{I}-10\log_{10}N+10\log_{10}(1-2\min_{k}f_{k}+\max_{k}f_{k}
\mathrm{)-10\log_{10}z-\log\min f}\label{eq:denumup}
\end{equation}
\end{widetext}

\noindent when there is no truncation due to coefficients being out
of bounds. 

The two bounds (equations \ref{eq:numup} and \ref{eq:denumup}) on
the SNR posit that for each scale, the SNR will decrease linearly
with slope one as a function of the input SNR, and will fall off
logarithmically per octave. That is, the actual data degrades at a rate of
-1.6 decibels per octave, and a linear fit to $10\log_{10}z$ would yield a
reduction of $0.92$dB per octave. Additional non-linear factors and other
factors not treated in this simple model are operative in the numerical 
simulations.

\section{Numerical Analysis\label{sec:Numerical-Analysis}}

\begin{figure*}
\includegraphics[width=1\textwidth]{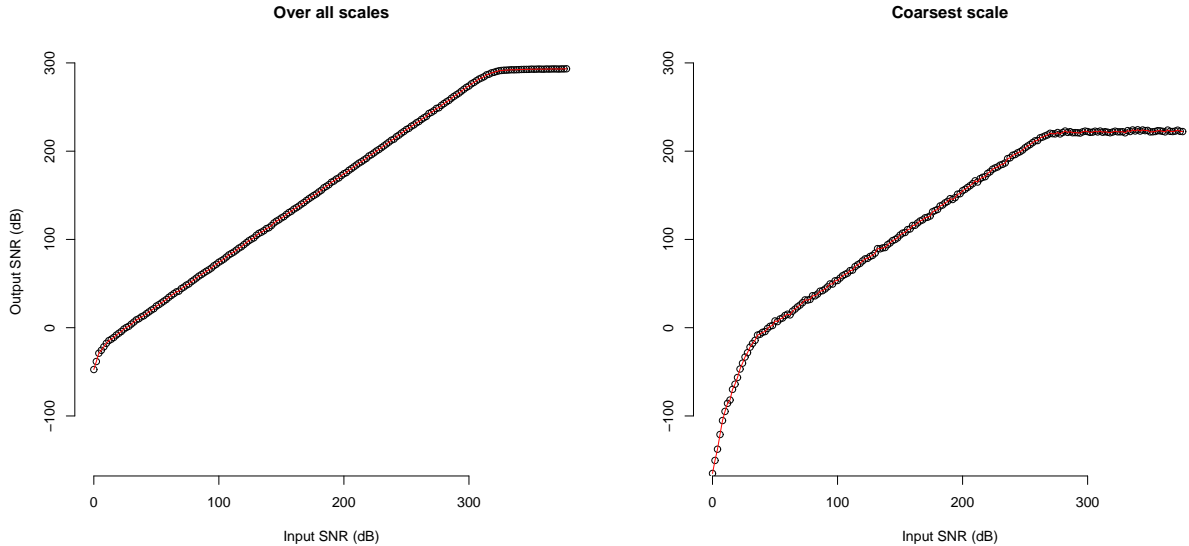} 
\caption{The relation between input and output SNR for the perturbed wavelet
filter over all wavelet scales ($2^{2}-2^{9}$, left panel) and for the coarse
scale ($2^{2}$, right panel) within double machine precision.\label{Flo:snr}}
\end{figure*}

To study truncation errors numerically, we generated a unit--power
white noise time series ($n=2^{14}$) (our results are qualitatively
the same for different classes of time series), downsampled the time
series by a factor of two using a FIR low pass filter of order $N=30$
(downsampling filter FIR1 \cite{signalprocessingtoolkit}), and chose
the difference of the low pass filter from the identity as our detail
coefficients. Coefficients used in a typical estimation procedure
depend on two integers: $m$, the number of times the data is downsampled
prior to estimation, and $k$, the number of octaves used in the estimation.
We used $m=7$ and $k=6$ in numerical analyses of the overall process
for definiteness. The SNR for each scale was also studied individually.
For error analysis, we perturbed the downsampling filter while systematically
varying the standard deviation, $\lambda_{i}$, of the noise $\epsilon_{i}$
(equation \ref{eq:perturbed_coef}) so that the input signal-to-noise
ratio $\left(\mathrm{SNR}_{I}\right)$ ranged from 380 to 0 decibels
(dB). For each input SNR, we chose 100 white noise time series, perturbed
the downsampling filter, and investigated the effect of the input
SNR for each time series to the output SNR of the filter given by 

\begin{equation}
\mathcal{O}(\mathrm{SNR}_{O})=10\log_{10}\dfrac{\sum_{ij}d_{ij}^{2}}{\sum_{ij}
(\tilde{d}_{ij}-d_{ij})^{2}}
\end{equation}

\noindent where $d_{ij}$ is the $i^{\mathrm{th}}$ detail coefficient
obtained at level $j$ by subtracting the low pass filtered value
from the coefficient prior to downsampling, and $\tilde{d}_{ij}$is
the same coefficient obtained from the perturbed filter (cf.
equation~\ref{eq:SNR_O}). 

The SNR of the output ($\mathrm{SNR}_{O}$) was approximately linearly
related to that of the input ($\mathrm{SNR}_{I}$) (Figure~\ref{Flo:snr},
left panel) with the equation

\noindent 
\begin{equation}
\mathcal{O}\left(\mathrm{SNR_{O}}\right)=-25.8+0.994\times\mathcal{O}
\left(\mathrm{SNR_{I}}\right)\label{eq:relation}
\end{equation}

\begin{figure}
\includegraphics[width=1\columnwidth]{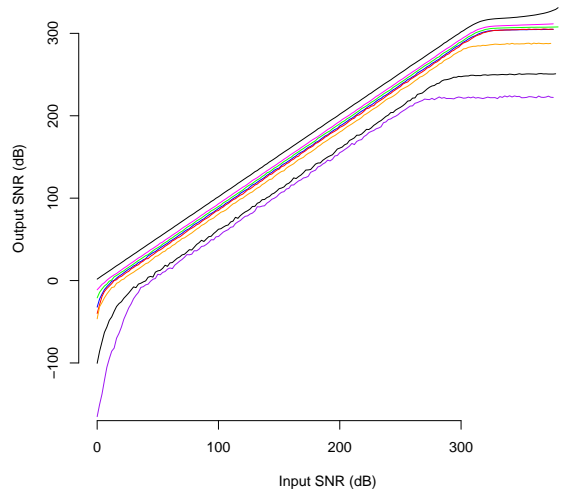}
\caption{Degradation of output SNR as a function of scale. Progressively lower
traces depict the relation for progressively coarser scales, from
$2^{13}$ to $2^{2}$.\label{fig:Degradation}}
\end{figure}

\noindent This relation shows that on average, a $10^{-4}$ error
in the estimated filter coefficients results in a $10^{-3}$ error
in the output. It should be noted, however, that this is a conservative
estimate for the overall output error. In fact, stringent constraints
may be necessary to attain sufficient accuracy at coarser scales because
of the degradation of output SNR (Figure~\ref{fig:Degradation}). 

\begin{figure}
\includegraphics[width=1\columnwidth]{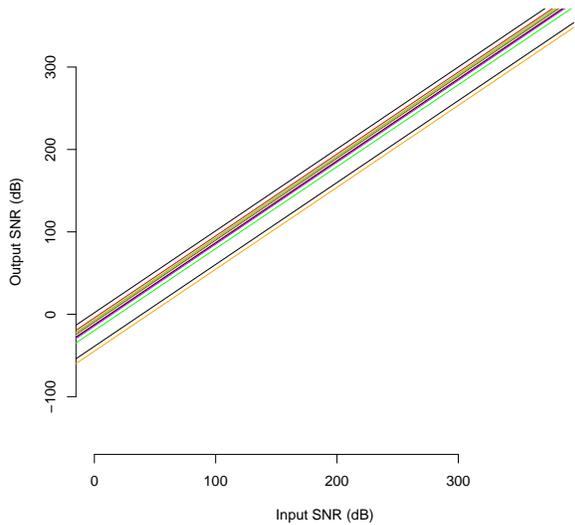}
\caption{The linear relation between input and output SNR for scales $2^{2}$
to $2^{13}$ in the $50-270$ dB input SNR range. Lower traces show
the relation for coarser scales.\label{Flo:scaleplot}}
\end{figure}

\begin{table}[h]
\caption{Growing intercept of the relation between input and output SNR with
coarser scales. See also Figure~\ref{Flo:scaleplot}.\label{Flo:sclint}}
\begin{tabular}{cc}
\textbf{Scale ($2^{n}$)} & \textbf{Intercept}\tabularnewline
\hline 
2  & -44.78\tabularnewline
3  & -39.86\tabularnewline
4  & -25.23\tabularnewline
5  & -14.59\tabularnewline
6  & -11.26\tabularnewline
7  & -9.87\tabularnewline
8  & -7.60\tabularnewline
9  & -4.61 \tabularnewline
\end{tabular}
\end{table}

Complete relation between $\mathcal{O}(\mathrm{SNR}_{O})$ and
$\mathcal{O}(\mathrm{SNR}_{I})$
for the range of double machine precision (Figure~\ref{Flo:snr},
right panel) shows that between approximately 50 to 270 dB input SNR,
output SNR is approximately a linear function (equation~\ref{eq:relation}).
In contrast, output SNR is flat when the input SNR is greater than
270 dB (where less than one bit of noise is added), and accelerates to 
$-\infty$ from 0 dB when the input SNR is below 50 dB. Moreover, although 
the output SNR is linearly related to that of the input with a slope of
1 for all scales ranging from $k=2^{2}$ to $k=2^{9}$ within the
range of $50-270$ dB input SNR (Figure~\ref{Flo:scaleplot}; cf.
equation~\ref{eq:relation}), the intercept of this relation degrades
at coarser scales (Table~\ref{Flo:sclint}). Therefore, coarser scales
are disproportionally affected. These results highlight that truncation
errors in wavelet coefficients can quickly reach to unacceptable levels.
For example, suppose that one requires the noise in the output of
the coarse scale wavelet coefficients to be less than $1$\% of the
output power. In this case, the noise power in the filter coefficients
must be 100 dB, or the filter coefficients must produce errors less
than approximately $5\times10^{-5}$. 

\begin{figure}
\includegraphics[width=1\columnwidth]{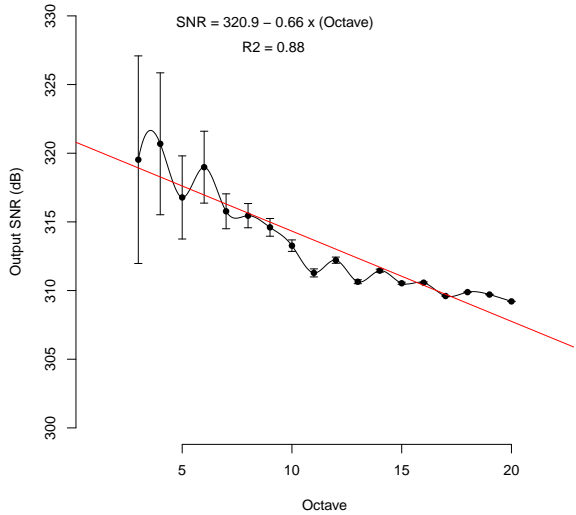}
\caption{An error analysis for the FFT of size $2^{8}$. The output SNR is
over $300$ dB, and the reduction in precision is about $0.85$ for
each additional octave. Vertical bars show 95\% confidence
intervals.\label{fig:fft}}
\end{figure}

To put this error amplification in perspective, consider simple filtering
via Fast Fourier Transform (FFT) of up to size $2^{20}$. In this case, SNR
of the signal degrades only by about 20 dB from its maximal value
of $306-325$ dB (Figure~\ref{fig:fft}). This relatively small loss
indicates that in practice, memory would be exhausted before the truncation
error becomes a problem. In contrast, the sub-band coding scheme reported
here (thus, most wavelet filters) require a significant length convolution
($N=30$) at each stage of the calculation. Thus, whereas the FFT
requires $\log_{2}N$ multiplications for each of the $N$ elements
inverted, wavelet filtering and sub-band coding schemes can require
$30^{\log_{2}N}$ calculations. This difference can be significant
at moderate input SNRs since repeated truncation errors may result
in a catastrophic loss of precision. 

\section{Conclusion}

Our results show that truncation error in the wavelet filter coefficients
due to quantization and round-off may amplify due to the the multiplicative
propagation, and can reach substantial levels. In cases where the
relative error in the wavelet coefficients is too large, the wavelet
tree $(k)$ grows too deep, or the wavelet filter $(N)$ is too long,
this propagation of truncation errors may render the results unreliable
especially at coarser scales. Note that although the sub-band analysis
presented here is not identical to downsampling using Coiflets or
Daubechies filters \cite{daubechies}, small errors
in the coefficients will still propagate in the latter case because
the propagation depends strongly on the length of the filter applied
prior to the downsampling. We expect errors with relatively large
D15 ($N=30$ coefficient Daubechies) filter to be comparable to the
case presented here.


\end{document}